\documentclass{revtex4}

\usepackage{graphicx}
\begin{document}
\bibliographystyle{revtex}


\title{PRECISION DETERMINATION OF $\rm V_{ub}$ AT AN $e^{+}e^{-}$ B FACTORY}



\author{Jik Lee} 
\email{jiklee@physics.purdue.edu}
\author{Ian Shipsey}
\email{shipsey@physics.purdue.edu}

\affiliation{Purdue University}


\date{\today}

\begin{abstract}
Current methods of 
determining $\rm V_{ub}$ are dominated by theoretical uncertainties. 
We present Monte Carlo simulations of three promising methods
of determining $\rm V_{ub}$ with small theoretical and experimental errors.
We find that with data samples of order 1,000 $fb^{-1}$ the 
B factories will attain combined experimental errors 
of a few \% on $\rm V_{ub}$, much smaller than the theoretical errors
associated with new inclusive methods. Lattice QCD offers the promise of rate 
calculations of exclusive semileptonic decays with errors of a 
few \%. A data sample of 
order 10,000 $fb^{-1}$, beyond the capabilities of the current B factories, 
may be required to achieve an experimental error on the exclusive rate 
comparable to the theoretical error.
\end{abstract}

\maketitle



\section{Current Methods}

CLEO was the first experiment to measure the ratio 
$|\rm V_{ub}|/|\rm V_{cb}|$ by detection of inclusive leptons
from semileptonic B meson decay beyond the kinematic endpoint for final
states containing charm mesons~\cite{cleo93}. 
Experimentally this is a clean measurement
but only 5-20\% of the leptons from  $b \to u \ell \nu$, depending 
on the model, populate the endpoint
region and so a large extrapolation, and hence a large theoretical 
uncertainty ($\sim$ 20\%) 
severely limits the precision of the measurement. 
However this method is still useful as a reality check of more precise 
methods.
More recently CLEO was able to reconstruct exclusive transitions using several 
analysis techniques that exploit the hermiticity of the CLEO detector to 
reconstruct the neutrino four vector~\cite{cleo96,cleo2000}.  Due to large 
backgrounds it has so far been 
found necessary to continue to work mostly in the lepton endpoint region. 
$|\rm V_{ub}|$ is extracted from the measured branching fraction, and the 
rate which is computed using a variety of theoretical methods. 
The theoretical 
error is large ($\sim $20\%)  and dominates the experimental errors
as the form factors governing 
the transitions are not derived from first 
principles, and the transitions are measured in only part of 
the phase space.  

\section{Future Methods}
There are two main theoretical approaches to improve the determination of 
$|\rm V_{ub}|$. 
The first of these is the operator product expansion (OPE) which is 
able to predict the inclusive $b \to u \ell \nu$ rate to $5-10\%$ within
experimentally tractable regions of phase space (either the region of
$m_{\rm had} < m_{\rm D}$ or the endpoint of the $q^2$ spectrum). 
The second theoretical approach is lattice QCD (LQCD). There has been 
a great deal of progress in LQCD in the last few years and within a few years
unquenched calculations of the rate for 
exclusive semileptonic decays will be available. The LQCD calculations may 
have an error of a few \% or better. 
Further discussion of the theoretical issues
in the extraction of a precision value of $|\rm V_{ub}|$ can be found 
in~\cite{Falk}~\cite{Zoltan}~\cite{lattice}.

\begin{figure}
\begin{center}
\includegraphics{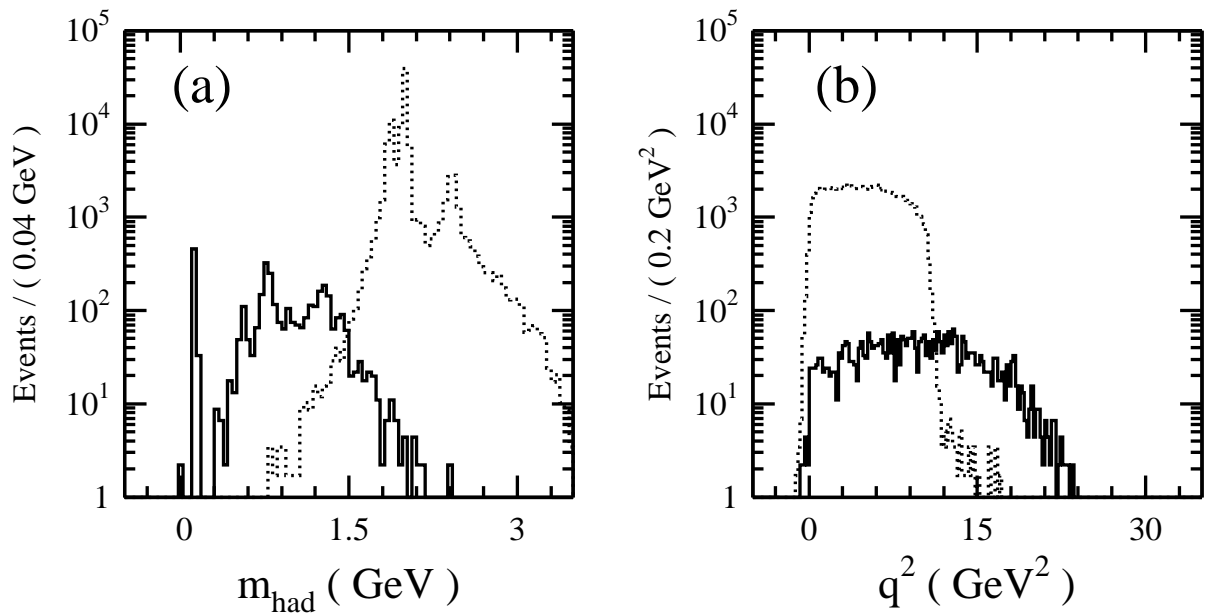}
\caption{(a) Hadronic mass ($m_{\rm had}$) distribution 
for 1000 $fb^{-1}$ data found with CLEO~III fast MC. 
The solid histogram is the $m_{\rm had}$ distribution of $b \to u \ell \nu$, 
and the dashed histogram is the $m_{\rm had}$ distribution 
of $b \to c \ell \nu$.
(b) $q^2$ distribution for 1000 $fb^{-1}$ data found with 
CLEO~III fast MC. 
The solid histogram is the $q^2$ distribution of $b \to u \ell \nu$, 
and the dashed histogram is the $q^2$ distribution of $b \to c \ell \nu$.}
\label{fig:mhadq2}
\end{center}
\end{figure}

The simulations reported here
are for a symmetric $e^+e^-$ B Factory. We use a fast (parametric)
Monte Carlo of the 
CLEO III detector called TRKSIM with the CLEO $\Upsilon(4S) \to B \bar B$ 
event generator. Efficiencies and $S/B$
for asymmetric B factories will differ slightly but the general 
conclusions are likely to be unchanged.   
Powerful suppression of the $b \to c \ell \nu$ background can be achieved
by full reconstruction of the companion $B$ decaying to
$B \to D^{(*)} (n\pi)$. As $B$ tagging has a relatively low 
efficiency ($2.85 \times 10^{-3}$)~\cite{efftag},
the technique was impractical for most analyses with pre B-factory
samples, but will be used extensively in future. 
We assume 1\% systematic error in lepton identification,
and 2\% systematic error in tracking in this paper.

\subsection{Inclusive hadronic mass spectrum}

The companion $B$ is reconstructed and the remainder of the event is then 
required
to have only one lepton, with momentum greater than 1.4 GeV, and missing 
mass consistent with a neutrino.
We select $b \to u \ell \nu$ events with $m_{\rm had} < m_{\rm D}$.
Figure~\ref{fig:mhadq2}(a) shows the simulated $m_{\rm had}$ distribution 
for an integrated luminosity of 1000 $fb^{-1}$.
Table~\ref{tab:mhadq2} shows the estimate of the statistical and systematic
error on $\rm V_{ub}$ measured with this method.

\begin{table*}[ht]
\begin{center}
\caption{Estimate of statistical and systematic
errors on $\rm V_{ub}$ measured with $m_{\rm had}$ and $q^2$ spectra,
where $S~(B)$ is the number of $b \to u~(c)~\ell \nu$ events.
The $m_{\rm had}$ method assumes a 10\% systematic error in $B$ 
for ${\cal L}_{int} = 100 fb^{-1}$.
It further assumes that this systematic error can be reduced
to 5\% for ${\cal L}_{int} = 100 fb^{-1}$
by vertexing, a better knowledge
of $D$ branching ratios, and a better knowledge of the form factors
in $B \to D/D^{*}/D^{**} \ell \nu$.
The $q^2$ method assumes a 100\% uncertainty in $B$ 
for ${\cal L}_{int} = 100 fb^{-1}$,
and a 20\% uncertainty in $B$ for ${\cal L}_{int} \ge 500 fb^{-1}$.
Note that errors in $\rm V_{ub}$ are experimental, not theoretical.}
\label{tab:mhadq2}
\vspace{0.1cm}
\begin{tabular}{cc|ccccc|ccccc} 
& & \multicolumn{5}{|c}{$m_{\rm had}$} &\multicolumn{5}{|c}{$q^2$} \\
\hline
year & ${\cal L}_{int} (fb^{-1})$ 
& $S$ & $B$ & \multicolumn{3}{c|}{$\delta \rm V_{ub}^{\rm expt.}$ (\%)} 
& $S$ & $B$ & \multicolumn{3}{c}{$\delta \rm V_{ub}^{\rm expt.}$ (\%)} \\
& & & & stat. & sys. & tot. & & & stat. & sys. & tot. \\
\hline
2002 &  100 &  335 &  127 & 3.2 & 2.2 & 3.9 &  127 &   7 & 4.6 & 3.0 & 5.5 \\
\hline					                                   
2005 &  500 & 1675 &  635 & 1.5 & 1.5 & 2.1 &  635 &  36 & 2.0 & 1.2 & 2.3 \\
\hline					                                   
2010 & 2000 & 6700 & 2540 & 0.7 & 1.5 & 1.7 & 2538 & 144 & 1.0 & 1.2 & 1.6 \\
\end{tabular}
\end{center}
\end{table*}

\begin{table*}[ht]
\begin{center}
\caption{Estimate of statistical and systematic
errors on $\rm V_{ub}$ measured
with the $B^0 \to \pi^- \ell^+ \nu$ decay.
We assume $S/B$ = 10/1, and a 10\% uncertainty in $B$ for this estimate.
The numbers in the parentheses are for $0.4<p_{\pi}< 0.8$ GeV/$c^2$. 
Note that errors in $\rm V_{ub}$ are experimental, not theoretical.}
\label{tab:lattice}
\vspace{0.2cm}
\begin{tabular}{ccccccc} 
year & ${\cal L}_{int} (fb^{-1})$ & $S$ & $B$ 
& \multicolumn{3}{c}{$\delta \rm V_{ub}^{\rm expt.}$ (\%)} \\
& & & & stat. & sys. & tot. \\
\hline
2008 & 1000 &  590(29) &  59(3) & 4.3(9.8) & 1.2 & 4.5(9.9) \\
\hline
 ?   &10000 & 5900(290)& 590(30)& 0.7(3.1) & 1.2 & 1.4(3.3) \\
\end{tabular}
\end{center}
\end{table*}

\subsection{Inclusive $q^2$ spectrum}

Using the inclusive $q^2$ endpoint results in a loss of statistics, 
but a gain
in theoretical certainty compared to the low $m_{\rm had}$ region.
The experimental advantage of this method
compared to the $m_{\rm had}$ method is that $S/B$ is more
favorable, therefore this method will be more attractive 
with very large data samples.
$B$ tagging and event selection are the same as in the previous method.
We select $b \to u \ell \nu$ events with $q^2 > 11.6$ GeV$^2$.
Figure~\ref{fig:mhadq2}(b) shows the simulated $q^2$ distribution  
for an integrated luminosity of 1000 $fb^{-1}$.
Table~\ref{tab:mhadq2} also shows the estimate of the statistical 
and systematic
experimental errors on $\rm V_{ub}$ obtained with this method.

\subsection{Exclusive decays with Lattice prediction} 

LQCD aims to predict the decay rate
of semileptonic decays such as $D \to \pi \ell \nu$ 
and  $B \to \pi \ell \nu$ to $\sim$ few \%,
which corresponds to $\delta \rm V_{ub}^{\rm stat.} 
\sim 1-2\%$~\cite{lattice}.
However many consistency checks will be required to 
demonstrate that the estimated lattice precision 
has been achieved. Here we outline one possible method.    

From an unquenched LQCD calculation of $f_D$ and 
a measurement of $D^+ \to \mu^+ \nu$
at a charm factory (for example  CLEO-C~\cite{cleo-c}) we obtain a precision 
direct measurement 
of $\rm {V_{cd}}$. Using this value of $\rm V_{cd}$, with an unquenched LQCD 
calculation of the rate and form factor 
shape of $D \to \pi \ell \nu$, we can make a direct test of the lattice
with a measurement 
of  $d\Gamma/dq^2(D \to \pi \ell \nu)$. This measurement is also
best done at a charm factory where
the kinematics at threshold cleanly separates signal from background.

If the lattice passes the above test, 
the second step is to compare the lattice prediction of the shape of 
$d\Gamma/dq^2(B \to \pi \ell \nu)$ to that of data at a B factory. 
If the shapes agree the third step
is to measure $\Gamma(B \to \pi \ell \nu)$ with data.
To cleanly isolate the signal we fully reconstruct the companion $B$ meson,
and select, for example, $B \to \pi \ell \nu$ events with the neutrino
reconstruction or consistency methods. 
Combining the measurement with the lattice prediction of
$\Gamma(B \to \pi \ell \nu)$, we extract $\rm V_{ub}$.
The theoretical error on $|\rm V_{ub}|$ may be as small as $1-2$\%.
Table~\ref{tab:lattice} shows the results of our simulation.
The numbers in the parentheses in Table~\ref{tab:lattice}
are obtained only using $\pi$'s in the range of $0.4<p_{\pi}< 0.8$ GeV/$c^2$,
where the systematic uncertainties involved in LQCD are well under
control, therefore LQCD is most reliable~\cite{lattice}.
To have a comparably small experimental error would require 
$\sim 10,000 fb^{-1}$.
Such large data samples are beyond the reach of existing B factories
where $\sim 2000 fb^{-1}$ are expected by the year 2010.

\section{Conclusion}

All possible theoretically clean measurements in the B sector 
are very important even if they are redundant within the standard model.
It is essential to pursue both CP violating and CP 
conserving measurements (i.e.,
$\rm V_{ub}$) to test the Standard Model and look for new physics.
Inclusive methods will achieve $\delta \rm V_{ub} 
\sim few \% (\rm experiment) \sim 5-10\% (\rm theory)$. The
$q^2$ endpoint is the method of choice.
The first test of CKM at the 10\% level will come from inclusive 
methods of determining $\rm V_{ub}$ and $\rm V_{cb}$, 
and the measurements of sin$2\beta$,
and $\rm V_{td}/\rm V_{ts}$.
If the lattice can reach the predicted accuracy ($1-2\%$), it will become
the method of choice for measurements of $\rm V_{ub}$ 
(and $\rm V_{cb}$) in the second decade of the 21st Century.
However the lattice must first be calibrated. A charm factory 
can provide unique and crucial tests of lattice predictions.
A $\sim 10,000 - 20,000 fb^{-1}$ data sample is required to attain
a total experimental error of $1-2\%$ on $\rm V_{ub}$ commensurate with 
the lattice error.

%
%

%
%


\bibliography{vub}

\begin{thebibliography}{8}
\expandafter\ifx\csname natexlab\endcsname\relax\def\natexlab#1{#1}\fi
\expandafter\ifx\csname bibnamefont\endcsname\relax
  \def\bibnamefont#1{#1}\fi
\expandafter\ifx\csname bibfnamefont\endcsname\relax
  \def\bibfnamefont#1{#1}\fi
\expandafter\ifx\csname citenamefont\endcsname\relax
  \def\citenamefont#1{#1}\fi
\expandafter\ifx\csname url\endcsname\relax
  \def\url#1{\texttt{#1}}\fi
\expandafter\ifx\csname urlprefix\endcsname\relax\def\urlprefix{URL }\fi
\providecommand{\bibinfo}[2]{#2}
\providecommand{\eprint}[2][]{\url{#2}}

\bibitem[{cle(1993)}]{cleo93}
\bibinfo{journal}{J. Bartelt {\it et al.}, Phys. Rev. Lett.}
  \textbf{\bibinfo{volume}{71}}, \bibinfo{pages}{4111} (\bibinfo{year}{1993}).

\bibitem[{cle(1996)}]{cleo96}
\bibinfo{journal}{J. P. Alexander {\it et al.}, Phys. Rev. Lett.}
  \textbf{\bibinfo{volume}{77}}, \bibinfo{pages}{5000} (\bibinfo{year}{1996}).

\bibitem[{cle(2000)}]{cleo2000}
\bibinfo{journal}{B. H. Behrens {\it et al.}, Phys. Rev. D}
  \textbf{\bibinfo{volume}{61}}, \bibinfo{pages}{052001}
  (\bibinfo{year}{2000}).

\bibitem[{\citenamefont{Falk}(2001)}]{Falk}
\bibinfo{author}{\bibfnamefont{A.}~\bibnamefont{Falk}},
  \bibinfo{journal}{Proceedings of the 4th International Conference on B
  Physics and CP violation}  (\bibinfo{year}{2001}).

\bibitem[{\citenamefont{Ligeti}(2000)}]{Zoltan}
\bibinfo{author}{\bibfnamefont{Z.}~\bibnamefont{Ligeti}},
  \bibinfo{journal}{''Prospects for $\rm V_{cb}$ and $\rm V_{ub}$'', Talk given
  at ''Beyond $10^{34}$ Physics at a Second Generation B Factory'',
  http://www.physics.purdue.edu/10E34/}  (\bibinfo{year}{2000}).

\bibitem[{lat(2001)}]{lattice}
\bibinfo{journal}{Aida X. El-Khadra {\it et al},hep-ph/0101023, and private
  communication}  (\bibinfo{year}{2001}).

\bibitem[{eff(1998)}]{efftag}
\bibinfo{journal}{The BaBar Physics Book, SLAC-R-504}  (\bibinfo{year}{1998}).

\bibitem[{cle(2001)}]{cleo-c}
\bibinfo{journal}{The CLEO-c Project Description,
  http://www.lns.cornell.edu/public/CLEO/spoke/CLEOc/}  (\bibinfo{year}{2001}).

\end{thebibliography}

\end{document}